\documentclass[11pt,a4paper]{article}

\usepackage{epsfig,amsmath,amssymb,cite}
\usepackage{hyperref}

\begin{document}

\title{Black hole solutions and thin shells in $N$-dimensional $F(R)$ gravity with a conformally invariant Maxwell field} 
\author{Ernesto F. Eiroa$^{1}$\thanks{e-mail: eiroa@iafe.uba.ar}, Griselda Figueroa-Aguirre$^{1}$\thanks{e-mail: gfigueroa@iafe.uba.ar}\\
{\small $^1$ Instituto de Astronom\'{\i}a y F\'{\i}sica del Espacio (IAFE, CONICET-UBA),}\\
{\small Casilla de Correo 67, Sucursal 28, 1428, Buenos Aires, Argentina}}
\date{}
\maketitle

\begin{abstract}

We present a generalization of the black hole solution with spherical symmetry already known in the literature for $N$-dimensional $F(R)$ gravity with a conformally invariant Maxwell field and constant scalar curvature $R$. This solution also includes a generalization of the one corresponding to general relativity as a special case. We introduce the formalism for the construction of a broad family of spherically symmetric thin shells in $F(R)$ theory. We use our generalized solution in order to provide examples of bubbles and thin layers of matter surrounding black holes. We analyze the stability of the constructions under perturbations preserving the symmetry, finding that stable configurations are possible for suitable values of the parameters. We show that the extension to higher dimensions does not alter the qualitative behavior of the thin shells found in four dimensions, with the main difference being a change of scale for the different values of $N$. 

\end{abstract}

\section{Introduction}\label{intro} 

Different theories of modified gravity have been proposed with the intention to explain some cosmological features, such as the early time inflation or the late accelerated expansion of the Universe, without the necessity of adding extra components like dark matter and dark energy. Among these theories, $F(R)$ gravity \cite{sofa} appears as one of the simplest modifications to general relativity, in which the scalar curvature $R$ in the Lagrangian density is replaced by an arbitrary function $F(R)$. In this theory, many solutions have been found in four dimensions; we can mention spherically symmetric black holes \cite{bhfr1,bhfr2,bhfr3} and traversable wormholes \cite{whfr1,whfr2}, among them. Higher dimensional spacetimes are mainly motivated by string theory; in this context, black hole solutions in general relativity \cite{BHRGndim1,BHRGndim2,BHRGndim3,BHRGndim4} as well as in $F(R)$ theories \cite{BHFRndim1,BHFRndim2,BHFRndim3} have been explored. Black holes in lower dimensionality have also been of interest, both in Einstein theory \cite{BTZ92,BHRG3dim1,BHRG3dim2} and in $F(R)$ gravity \cite{BHFR3dim}.

In general relativity, the formalism that allows the construction of a new spacetime by joining different geometries across a hypersurface was developed by Darmois and Israel \cite{daris}. The technique provides the conditions that thin layers of matter have to satisfy for a proper matching, allowing to study their dynamics from the analysis of the energy-momentum tensor on the hypersurface. In particular, for highly symmetric scenarios, the stability of the configurations under perturbations that preserve the symmetry can be studied analytically. This formalism has been used in many scenarios due to its simplicity and flexibility; examples in four dimensions can be found in models of gravastars \cite{gravstar}, wormholes \cite{whrg}, bubbles, and thin layers of matter surrounding black holes \cite{shrg}. There are also some studies in which the junction conditions are used to build wormholes and thin shells of matter in $N$ dimensions \cite{GRddim}.

The Darmois–Israel formalism has been extended in recent years to $F(R)$ theories \cite{dss,js}, displaying junction conditions of a more restrictive nature than those in general relativity, in them the continuity of the trace of the second fundamental form at the matching hypersurface is always required. Non-quadratic $F(R)$ gravity also demands the continuity of the scalar curvature across it, while quadratic $F(R)$ allows its discontinuity; in this case, extra contributions appear besides the energy-momentum tensor, in order to guarantee local conservation \cite{js}. Different physical objects have been studied within $F(R)$ gravity in four dimensions by using this technique \cite{TSWHinFR,tsFR} and also in lower dimensionality \cite{tsFR3d}. However, works considering thin shells in higher dimensional $F(R)$ gravity have been scarce. 

In this article, we obtain a generalization of the known spherically symmetric solution \cite{BHFRndim2} for $N$-dimensional $F(R)$ gravity with a conformally invariant Maxwell field; as a byproduct, we also derive a generalization of the corresponding spacetime previously found \cite{BHRGndim1} within general relativity. We subsequently study thin shells of matter within this context. In Sect. \ref{sol-fr}, we present the generalized black hole solution with a constant scalar curvature. In Sect. \ref{construction}, we introduce the general formalism for the construction of a class of $N$-dimensional spherical thin shells in $F(R)$ gravity with constant scalar curvature. In Sect. \ref{stability}, we develop the stability analysis of the static configurations under radial perturbations. In Sect. \ref{examples}, we show examples of bubbles and thin layers of matter surrounding black holes, in which we use our generalized solution. Finally, we present a summary in Sect. \ref{summary}. We adopt units such that $c=G_N=1$.

\section{Black hole solutions with charge in $F(R)$ gravity}\label{sol-fr}

In $F(R)$ gravity coupled to a power law nonlinear electrodynamics, we adopt the $N$-dimensional action 
\begin{equation}
I =\frac{1}{16\pi }\int d^{N}x\sqrt{-g}\left(R+ f(R) -\alpha \varepsilon |\mathcal{F}| ^{s}\right) ,
\label{Action}
\end{equation}
where $R+ f(R)=F(R)$ is the gravitational Lagrangian (the first term corresponds to general relativity) and $\mathcal{F} = \mathcal{F}_{\alpha\beta}\mathcal{F}^{\alpha\beta}$ denotes the Maxwell invariant, with $\mathcal{F}_{\mu \nu }=\partial _{\mu }\mathcal{A}_{\nu }-\partial _{\nu }\mathcal{A}_{\mu }$ the electromagnetic tensor field defined in terms of the gauge potential $\mathcal{A}_{\mu }$. The constant $s\neq 1/2$ is a positive nonlinearity parameter, $\varepsilon = \mathrm{sign} (\mathcal{F})$, and the constant $\alpha$ can be taken as $1$ or $-1$ with a suitable choice of electromagnetic units, without losing generality. When $s=1$, the standard linear Maxwell field term in the action is recovered, with the usual sign for $\alpha =1$ or the opposite sign for $\alpha =-1$. The field equations resulting from this action in the metric formalism read
\begin{equation}
R_{\mu\nu}(1+f'(R)) - \frac{1}{2}g_{\mu\nu}(R+f(R))+
(g_{\mu\nu}\nabla_\gamma \nabla^\gamma -\nabla_\mu \nabla_\nu)f'(R)=8\pi T_{\mu\nu},
\label{field_eqns}
\end{equation}
\begin{equation}
\nabla_\mu \left(\mathcal{F}^{\mu\nu}|\mathcal{F}|^{s-1}\right)=\frac{1}{\sqrt{-g}}\partial_\mu \left(\sqrt{-g}\mathcal{F}^{\mu\nu}|\mathcal{F}|^{s-1}\right)=0.
\label{em_eqns}
\end{equation}
The energy-momentum tensor associated with the electromagnetic field has the form 
\begin{equation}
T_{\mu\nu}=\frac{\alpha}{4\pi}\left[ s\mathcal{F}_{\mu \gamma}\mathcal{F}_{\nu}^{\;\;\gamma}|\mathcal{F}|^{s-1}-\frac{1}{4} g_{\mu\nu}\varepsilon |\mathcal{F}|^s\right],\end{equation}
so its trace is
\begin{equation}
T^{\mu}_{\;\;\mu}=\frac{\alpha \varepsilon}{4\pi}|\mathcal{F}|^s \left(s-\frac{N}{4}\right).
\label{Tem}
\end{equation}
The case $s = N/4$ gives a traceless $T_{\mu\nu}$, corresponding to a conformally invariant Maxwell field as a source; then, we adopt $s=N/4$ in what follows. When $N=4$ results in $s=1$, so in this case we recover the linear Maxwell field; otherwise, the electromagnetic field is nonlinear. We adopt the spherically symmetric ansatz
\begin{equation}
ds^2 = -A(r)dt^2 + A(r)^{-1}dr^2 +r^2d\Omega^{N-2} ,
\label{metric-N-sphe}
\end{equation}
with $t$ the time coordinate, $r>0$ the radial coordinate, $0\le \theta _i\le \pi$ ($1\le i \le N-3$) and $0\le \theta _{N-2} < 2 \pi $ the angular coordinates, so that
\begin{equation}
d\Omega^2_{N-2}= d\theta^2_1+\sideset{}{}\sum_{i=2}^{N-2} \sideset{}{ }\prod_{j=1}^{i-1} \sin^2 \theta _j d\theta_i^2.
\end{equation}
We consider a radial purely electric field $E(r)=F_{tr}$; then, $\mathcal{F}=-2(\mathcal{F}_{tr})^2$ is negative and $\varepsilon=-1$ in this case. By substituting all this in Eq. (\ref{em_eqns}), we find that $F_{tr}=Q/r^2$, with $Q$ the electric charge, is independent of the dimension $N$. After some calculations we obtain, for a constant scalar curvature $R_0$, that Eq. (\ref{field_eqns}) admits a solution where the metric function reads
\begin{equation}
A(r)= -M + \frac{\alpha   |Q|^{3/2} }{2^{1/4}(1+f'(R_0)) r}-\frac{R_0 r^2}{6} \qquad \mathrm{if} \; N=3
\label{Ametric3d}
\end{equation}
and
\begin{equation}
A(r)=   1- \frac{2 M}{ r^{N-3}}+ \frac{\alpha  2^{N/4} |Q|^{N/2} }{2(1+f'(R_0)) r^{N-2}}-\frac{R_0 r^2}{(N-1) N}  \qquad \mathrm{if} \; N \ge 4,
\label{AmetricNd}
\end{equation}
where the constant $M$ represents the mass. The spacetime has a curvature singularity at $r=0$ since the Kretschmann scalar diverges at this point. The trace of the field equations gives 
\begin{equation}
R_0 \left(1+f'(R_0)\right)-\frac{N}{2} \left(R_0+f(R_0)\right)=0 ,
\label{trace1}
\end{equation}
which allows us to define the effective cosmological constant $\Lambda_e$ 
\begin{equation}
R_0=\frac{Nf(R_0)}{2f'(R_0)+2-N} \equiv \frac{2N}{N-2}\Lambda_e .
\label{trace2}
\end{equation}
On the other hand, the traceless energy-momentum tensor of the electromagnetic field (\ref{Tem}) has an energy density in the associated orthonormal frame given by $T_{\hat{t}\hat{t}}$, which in our case results
\begin{equation}
T_{\hat{t}\hat{t}}=\frac{\alpha}{32\pi}(2F_{tr}^2)^{N/4}(N-2)=\frac{\alpha}{32\pi}2^{N/4}|Q|^{N/2}(N-2);
\end{equation}
this quantity is positive for $N\ge3$ if and only if $\alpha >0$. Then, there are two branches in our solution, but the one with $\alpha <0$ has a negative energy density,  so in what follows we take $\alpha =1$. 

A particular case of interest results from taking $f(R)=-2\Lambda$, with $\Lambda$ the cosmological constant\footnote{Note that from Eq. (\ref{trace2}) one obtains that $\Lambda_e = \Lambda$.}, corresponding to general relativity. In this scenario, if $N=3$ from Eq. (\ref{Ametric3d}), we retrieve the geometry found in Refs. \cite{BHRG3dim1,BHRG3dim2}; for $N=4$, we recover from Eq. (\ref{AmetricNd}) the Reissner-Nordstr\"om ($\Lambda =0$), the Reissner-Nordstr\"om-de Sitter ($\Lambda >0$), and the Reissner-Nordstr\"om-anti-de Sitter ($\Lambda <0$) solutions; while for $N>4$ we obtain a generalization to arbitrary $N$ of the geometry introduced in Ref. \cite{BHRGndim1}, in which the dimension $N$ of the spacetime is restricted to a set of particular values, that is, $N=4+4p$, with $p\in \mathbb{N}$. The introduction of the absolute value of the Maxwell invariant $\mathcal{F}$ in our Lagrangian is the key point to overcome this restriction on $N$; in our treatment, the sign of $\mathcal{F}$ is taken into account by the presence of the factor $\varepsilon$ outside the power law exponent $s$. The absolute value of $\mathcal{F}$ in the Lagrangian has been previously used within general relativity by other authors \cite{BHRG3dim2,BHRGndim3}, but the extra factor $\varepsilon$ adopted in our work seems to be novel. Note the different nature of the constants $\alpha$ and $\varepsilon$: the first one represents the coupling between gravity and electrodynamics, while the second one only depends on the characteristics of the electromagnetic field (in our case purely electric).

Returning to the general $F(R)$ theory, for $N=3$, the existence of the black hole requires $R_0<0$, that is, an anti-de Sitter asymptotics. When $N=3$ and $\alpha =-1$, we easily recover the spacetime\footnote{Note that in this case is $T_{\hat{t}\hat{t}}<0$, but our solution also allows to adopt $\alpha =1$, so that $T_{\hat{t}\hat{t}}>0$.} found in Ref. \cite{BHFR3dim}. If $N \ge 4$, the geometry is asymptotically anti-de Sitter, Minkowski, or de Sitter, depending on $R_0 <0$, $R_0 =0$, or $R_0>0$, respectively. For $R_0\le 0$, when $|Q|$ is small enough the spacetime has an event horizon with a radius given by the largest real positive solution of the equation $A(r)=0$, while for large $|Q|$, there is a naked singularity. For $R_0>0$, the largest real positive solution of the equation $A(r)=0$ gives the radius of the cosmological horizon, while for small enough $|Q|$, the second largest one gives the radius of the event horizon; again for large $|Q|$ the singularity is naked.   When $N=4$ and $\alpha =1$, it is straightforward to see that the geometry obtained in Ref. \cite{bhfr2} is recovered. For $N>4$, our solution generalizes the one found in Ref. \cite{BHFRndim2}, for which the same considerations stated above in the case of Ref. \cite{BHRGndim1} also applies, i.e., there are no restrictions on the possible values of $N$.

\section{Thin shell construction} \label{construction}

In this section, we describe the formalism for spherical thin shells in $N$-dimensional $F(R)$ gravity. The manifold $\mathcal{M}$ is crafted as the union of two different ones $\mathcal{M}_1$ and $\mathcal{M}_2$, each of them with a constant scalar curvature $R_{1,2}$. We paste these manifolds on a spherical hypersurface $\Sigma$ with radius $a$, where there is a thin layer of matter. The original manifolds $\mathcal{M}_{1,2}$ are described by two different spherically symmetric metrics of the form shown in Eq. (\ref{metric-N-sphe}), with generic metric functions $A_{1,2} (r_{1,2})$ and coordinates $(t_{1,2},r_{1,2},\theta_1, ... ,\theta_{N-2})$. We define the inner manifold  $\mathcal{M}_1$ by taking $0\le r_1 \leq a$ and the outer manifold $\mathcal{M}_2$ by $r_2\geq a$, and we paste them at the hypersurface $\Sigma $ to create a new manifold $\mathcal{M}=\mathcal{M}_1 \cup \mathcal{M}_2$. We have mutually identified the angular coordinates of both original spacetimes since our construction takes into consideration the spherical symmetry. A new global radial coordinate can be defined as $r\in [0,+\infty)$ by identifying $r$ with $r_1$ in $\mathcal{M}_1$ and with  $r_2$ in $\mathcal{M}_2$. The global coordinates are denoted by $X^{\alpha }_{1,2} = (t_{1,2},r,\theta_1, ... ,\theta_{N-2})$ while the coordinates on the hypersurface $\Sigma $ are $\xi ^{i}=(\tau ,\theta_1, ... ,\theta_{N-2})$, with $\tau $ the proper time there. In order to analyze the dynamics of this hypersurface, we take its radius as $a(\tau)$. The proper time should be the same when obtained from each side of $\Sigma $. Then, the different coordinate times of the original manifolds can be related with $\tau $ via
\begin{equation}
\frac{dt_{1,2}}{d\tau} = \frac{\sqrt{A_{1,2}(a) + \dot{a} ^2}}{A_{1,2}(a)},
\label{tau}
\end{equation}
where the signs are determined by choosing $t_{1,2}$ and $\tau$ to run into the direction of the future. We also denote $\dot{a}$ as the time derivative of $a$. To continue with our construction, we need to calculate the first fundamental form for each original manifold 
\begin{equation}
h^{1,2}_{ij}= \left. g^{1,2}_{\mu\nu}\frac{\partial X^{\mu}_{1,2}}{\partial\xi^{i}}\frac{\partial X^{\nu}_{1,2}}{\partial\xi^{j}}\right| _{\Sigma }
\end{equation}
and the second fundamental form, also named extrinsic curvature, 
\begin{equation}
K_{ij}^{1,2 }=-n_{\gamma }^{1,2 }\left. \left( \frac{\partial ^{2}X^{\gamma
}_{1,2} } {\partial \xi ^{i}\partial \xi ^{j}}+\Gamma _{\alpha \beta }^{\gamma }
\frac{ \partial X^{\alpha }_{1,2}}{\partial \xi ^{i}}\frac{\partial X^{\beta }_{1,2}}{
\partial \xi ^{j}}\right) \right| _{\Sigma },
\label{sff}
\end{equation}
where the unit normals ($n^{\gamma }n_{\gamma }=1$) can be obtained from
\begin{equation}
n_{\gamma }^{1,2 }=\left\{ \left. \left| g^{\alpha \beta }_{1,2}\frac{\partial G}{\partial
X^{\alpha }_{1,2}}\frac{\partial G}{\partial X^{\beta }_{1,2}}\right| ^{-1/2}
\frac{\partial G}{\partial X^{\gamma }_{1,2}} \right\} \right| _{\Sigma },
\end{equation}
with $G(r)\equiv r-a =0$ on $\Sigma $. We will work in the orthonormal basis at the hypersurface $\Sigma$ defined by
\begin{equation*}
e_{\hat{\tau}}=e_{\tau }, \qquad e_{\hat{\theta}_1}=a^{-1}e_{\theta_1 }, \qquad e_{\hat{\theta}_i}=\left( a\prod_{j=1}^{i-1} \sin \theta _j \right)^{-1} e_{\theta_i } \quad \mathrm{if} \quad 2\le i \le N-2 ,
\end{equation*}
because it facilitates the physical interpretation of the results. Once we have all these elements well defined, we can calculate them for the metric given by Eq. (\ref{metric-N-sphe}). In this case, by using Eq. (\ref{tau}), we find that the first fundamental form is \begin{equation}
h^{1,2}_{\hat{\imath}\hat{\jmath}}= \mathrm{diag}(-1,1,...,1), 
\label{h-metric}
\end{equation}
the unit normals are given by 
\begin{equation} 
n_{\gamma }^{1,2}= \left(-\dot{a},\frac{\sqrt{A_{1,2}(a)+\dot{a}^2}}{A_{1,2}(a)},0,...,0 \right),
\end{equation}
while the non-null elements of the second fundamental form read
\begin{equation} 
K_{\hat{\tau}\hat{\tau}}^{1,2 }=-\frac{A '_{1,2}(a)+2\ddot{a}}{2\sqrt{A_{1,2}(a)+\dot{a}^2}},
\label{e5}
\end{equation}
and
\begin{equation} 
K_{\hat{\theta}_i\hat{\theta}_i}^{1,2}=\frac{1}{a}\sqrt{A_{1,2} (a) +\dot{a}^2},
\label{e4}
\end{equation}
where the prime on $A_{1,2}(r)$ represents the derivative with respect to $r$. It is also worth defining that a prime on $F(R)$ corresponds the derivative with respect to the curvature scalar $R$ and the jump  of any quantity $\Upsilon $ across $\Sigma$ is denoted by $[\Upsilon ]\equiv (\Upsilon ^{2}-\Upsilon  ^{1})|_\Sigma $.

According to the junction formalism in $F(R)$ gravity \cite{js}, there is a set of conditions that should be satisfied for a proper matching at the hypersurface  $\Sigma $. The continuity of the first fundamental form is always required, that is,  $[h_{\mu \nu}]=0$. Due to the nature of our construction, this is satisfied automatically, as we can see from Eq. (\ref{h-metric}). It also demands the continuity of the trace of the second fundamental form, that is, $[K^{\mu}_{\;\; \mu}]=0$, which can be written, after manipulating Eqs. (\ref{e5}) and (\ref{e4}), as
\begin{equation} 
\frac{2\ddot{a}+ A_{2}'(a)}{2\sqrt{A_{2}(a)+\dot{a}^2}}-\frac{2\ddot{a}+ A_{1}'(a)}{2\sqrt{A_{1}(a)+\dot{a}^2}}+\frac{(N-2)}{a}\left(\sqrt{A_{2}(a)+\dot{a}^2}-\sqrt{A_{1}(a)+\dot{a}^2}\right)=0.
\label{CondGen}
\end{equation}
The formalism divides into two branches that, depending on the third derivative $F'''(R)$, in one case demands an extra condition \cite{js} for a proper matching at $\Sigma$.

\subsection{Case $F'''(R) \neq 0$}

When  $F'''(R) \neq 0$, the continuity of $R$ across $\Sigma$ is also required \cite{js}, that is, $[R]=0$. In this case, the field equations at $\Sigma$ \cite{js} read
\begin{equation} 
\kappa S_{\mu \nu}=-F'(R)[K_{\mu \nu}]+ F''(R)[\eta^\gamma \nabla_\gamma R]  h_{\mu \nu}, \;\;\;\; n^{\mu}S_{\mu\nu}=0,
\label{LancGen}
\end{equation}
with $\kappa =8\pi $ and $S_{\mu \nu}$ the energy--momentum tensor at $\Sigma$. Since $[R]=0$, the values of the scalar curvature at both sides of $\Sigma$ should be the same, i.e., $R_1=R_2=R_0$, and these equations for constant $R_0$ simplify to give
\begin{equation} 
\kappa S_{\mu \nu}=-F'(R_0)[K_{\mu \nu}], \;\;\;\; n^{\mu}S_{\mu\nu}=0.
\label{LanczosGen}
\end{equation}

Using the orthonormal basis, the energy--momentum tensor takes the diagonal form $S_{_{\hat{\imath}\hat{\jmath} }}=\mathrm{diag}(\sigma ,p,...,p)$, with $\sigma$ the hypersurface energy density and $p \equiv p_{\hat{\theta}_i}$ ($1 \le i \le N-2$) the transverse pressure. Then, we obtain
\begin{equation} 
\sigma= -\frac{F'(R_0)}{\kappa }\left( -\frac{2\ddot{a}+A_{2}'(a)}{2\sqrt{A_{2}(a)+\dot{a}^2}}+\frac{2\ddot{a}+A_{1}'(a)}{2\sqrt{A_{1}(a)+\dot{a}^2}}\right)
\label{e9}
\end{equation}
and
\begin{equation}
p=\frac{- F'(R_0)}{a\kappa }\left( \sqrt{A_{2}(a)+\dot{a}^2}-\sqrt{A_{1}(a)+\dot{a}^2}\right).
\label{e10}
\end{equation}
By algebraically working with Eqs. (\ref{CondGen}), (\ref{e9}), and (\ref{e10}), we can see that the equation of state has the form $\sigma -(N-2)p=0$.

\subsection{Case $F'''(R) = 0$}

When $F'''(R) = 0$, we are in the presence of quadratic $F(R)$ theory, i.e., $F(R)=R-2\Lambda+\gamma R^2$ and therefore $F'(R)= 1+2\gamma R$. In this case, the condition of the continuity of the scalar curvature at the matching hypersurface is no longer required  \cite{js}. The field equations at $\Sigma$ \cite{js} have the form
\begin{equation}
\kappa S_{\mu \nu} =-[K_{\mu\nu}]+2\gamma \left( [n^{\gamma }\nabla_{\gamma}R] h_{\mu\nu}-[RK_{\mu\nu}] \right), \;\;\;\; n^{\mu}S_{\mu\nu}=0;
\label{LancQuad}
\end{equation}
which, for constant values $R_1$ and $R_2$ of the scalar curvature at the sides of $\Sigma$, reduce to
\begin{equation}
\kappa S_{\mu \nu} =-[K_{\mu\nu}]-2\gamma[RK_{\mu\nu}], \;\;\;\; n^{\mu}S_{\mu\nu}=0.
\label{LanczosQuad}
\end{equation}
The presence of three extra contributions \cite{js} is needed to guarantee that the energy--momentum tensor is divergence--free and, therefore, locally conserved. These contributions are: an external scalar pressure or tension
\begin{equation}
\kappa\mathcal{T}=2\gamma [R] K^\gamma{}_\gamma ,
\label{Tg}
\end{equation}
an external energy flux vector
\begin{equation}
\kappa\mathcal{T}_\mu=-2\gamma \bar{\nabla}_\mu[R]=0,  \qquad  n^{\mu}\mathcal{T}_\mu=0,
\label{Tmu}
\end{equation}
with $\bar{\nabla }$ the intrinsic covariant derivative on $\Sigma$, and a two-covariant symmetric tensor distribution 
\begin{equation}
\kappa \mathcal{T}_{\mu \nu}=\nabla_{\beta } \left( 2\gamma [R] h_{\mu \nu } n^{\beta } \delta ^{\Sigma }\right),
\label{dlay1}
\end{equation}
with $\delta ^{\Sigma }$ the Dirac delta on $\Sigma $. This last contribution has resemblance with classical dipole distributions. Then, in this case, the singular part at the shell of the energy--momentum tensor reads $(S_{\mu \nu} + \mathcal{T}_\mu n_\nu + \mathcal{T}_\nu n_\mu + \mathcal{T} n_\mu n_\nu )\delta ^{\Sigma } + \mathcal{T}_{\mu \nu}$; for further details, we suggest to read Ref. \cite{js}. Since we are adopting constant values of the scalar curvature at both sides of the shell, the external energy flux vector is always null, i.e., $\mathcal{T}_\mu=0$. 

As it was stated in the previous case, working in the orthonormal basis allows us to have a diagonal energy--momentum tensor $S_{_{\hat{\imath}\hat{\jmath} }}=\mathrm{diag}(\sigma ,p,...,p)$, in which the hypersurface energy density and the pressure are given by
\begin{equation} 
\sigma= \frac{1+2\gamma R_2}{\kappa }\left( \frac{2\ddot{a}+A_{2}'(a)}{2\sqrt{A_{2}(a)+\dot{a}^2}} \right)- \frac{1+2\gamma R_1}{\kappa }\left( \frac{2\ddot{a}+A_{1}'(a)}{2\sqrt{A_{1}(a)+\dot{a}^2}}\right),
\label{e9Rdif}
\end{equation}
and
\begin{equation}
p= -\frac{1+2\gamma R_2}{\kappa }\left(\frac{\sqrt{A_{2}(a)+\dot{a}^2}}{a}\right)+ \frac{1+2\gamma R_1}{\kappa }\left(\frac{\sqrt{A_{1}(a)+\dot{a}^2}}{a}\right).
\label{e10Rdif}
\end{equation}
By using Eqs. (\ref{CondGen}) and (\ref{Tg}), the external scalar pressure or tension can be written as
\begin{eqnarray}
\mathcal{T} &=& \frac{2\gamma R_2}{\kappa }\left( \frac{2\ddot{a}+A_{2}'(a)}{2\sqrt{A_{2}(a)+\dot{a}^2}} + (N-2) \frac{\sqrt{A_{2}(a)+\dot{a}^2}}{a} \right) \nonumber \\
& & - \frac{2\gamma R_1}{\kappa }\left( \frac{2\ddot{a}+A_{1}'(a)}{2\sqrt{A_{1}(a)+\dot{a}^2}}+ (N-2)\frac{\sqrt{A_{1}(a)+\dot{a}^2}}{a}\right),
\label{e11Rdif}
\end{eqnarray} 
while the double layer tensor distribution $\mathcal{T}_{\hat{\imath}\hat{\jmath}}$ is proportional to $2 \gamma [R] h_{\hat{\imath}\hat{\jmath}} /\kappa$. From Eqs. (\ref{e9Rdif}), (\ref{e10Rdif}), and (\ref{e11Rdif}), with the help of Eq. (\ref{CondGen}), we find that the equation of state is $\sigma - (N-2) p=\mathcal{T}$ in this case. 

In both scenarios, we say that the matter at the shell is normal when satisfies the weak energy condition (WEC), that is,  $\sigma \geq 0$ and $\sigma + p \geq 0$; otherwise, it is exotic. On the other hand, within $F(R)$ gravity, the inequality $F'(R)>0$ is required in order to have a positive effective Newton constant $G_{eff} = G/F'(R_0)$, and therefore, prevent the graviton from being a ghost \cite{bronnikov}.

\section{Stability analysis} \label{stability}

Let us consider the static configurations, which has to satisfy the static version of Eq. (\ref{CondGen}), that is,
\begin{equation} 
\frac{ A_{2}'(a_0)}{2\sqrt{A_{2}(a_0)}}-\frac{ A_{1}'(a_0)}{2\sqrt{A_{1}(a_0)}}+\frac{(N-2)}{a_0}\left(\sqrt{A_{2}(a_0)}-\sqrt{A_{1}(a_0)}\right)=0,
\label{CondEstatico}
\end{equation}
with $a_0$ denoting the constant radius of the shell. Considering the matter content at the shell, in the case that $F'''(R)\neq 0$, the energy density $\sigma _0$ and the pressure $p _0$ at $\Sigma$ take the form
\begin{equation} 
\sigma_0= -\frac{F'(R_0)}{\kappa }\left( -\frac{A_{2}'(a_0)}{2\sqrt{A_{2}(a_0)}}+\frac{A_{1}'(a_0)}{2\sqrt{A_{1}(a_0)}}\right)
\label{e13}
\end{equation}
and
\begin{equation}
p_0=\frac{- F'(R_0)}{a_0\kappa }\left( \sqrt{A_{2}(a_0)}-\sqrt{A_{1}(a_0)}\right),
\label{e14}
\end{equation}
which satisfy the equation of state is $\sigma_0 - (N-2) p_0=0$. When $F'''(R)= 0$, we obtain that the expressions for the energy density and the pressure read
\begin{equation} 
\sigma_0= \frac{1+2\gamma R_2}{\kappa }\left( \frac{A_{2}'(a_0)}{2\sqrt{A_{2}(a_0)}} \right)- \frac{1+2\gamma R_1}{\kappa }\left( \frac{A_{1}'(a_0)}{2\sqrt{A_{1}(a_0)}}\right)
\label{e13DifR}
\end{equation}
and
\begin{equation}
p_0= -\frac{1+2\gamma R_2}{\kappa }\left(\frac{\sqrt{A_{2}(a_0)}}{a_0}\right)+ \frac{1+2\gamma R_1}{\kappa }\left(\frac{\sqrt{A_{1}(a_0)}}{a_0}\right).
\label{e14DifR}
\end{equation}
The external scalar pressure or tension results
\begin{equation}
\mathcal{T}_0=\frac{2\gamma R_2}{\kappa }\left( \frac{A_{2}'(a_0)}{2\sqrt{A_{2}(a_0)}} + (N-2) \frac{\sqrt{A_{2}(a_0)}}{a_0} \right)- \frac{2\gamma R_1}{\kappa }\left( \frac{A_{1}'(a_0)}{2\sqrt{A_{1}(a_0)}}+ (N-2)\frac{\sqrt{A_{1}(a_0)}}{a_0}\right),
\label{e16DifR}
\end{equation}
and the equation of state now is $\sigma_0 - (N-2) p_0=\mathcal{T}_0 $. The other non-null extra contribution that appears in this case is the double layer tensor distribution $\mathcal{T}^{(0)}_{\hat{\imath}\hat{\jmath}}$, proportional to $2 \gamma [R] h_{\hat{\imath}\hat{\jmath}} /\kappa$.

The stability analysis of the static shells under radial perturbations is facilitated by Eq. (\ref{CondGen}). If we use $\ddot{a}= (1/2)d(\dot{a}^2)/da$ and define $z=\sqrt{A_{2}(a)+\dot{a}^2}-\sqrt{A_{1}(a)+\dot{a}^2}$, we can rewrite this equation in the form
\begin{equation}
az'(a)+2z(a)=0.
\label{CondGen_u}
\end{equation}
We can solve the equation above to obtain an expression for $\dot{a}^{2}$ in terms of $a$, that is,
\begin{equation}
\dot{a}^{2}=-V(a),
\label{condicionPot}
\end{equation}
where 
\begin{equation}
V(a)= -\frac{a_{0}^{2N-4}\left(\sqrt{A_{2}(a_{0})}-\sqrt{A_{1}(a_{0})}\right)^2}{4a^{2N-4}} +\frac{A_{1}(a)+A_{2}(a)}{2} -\frac{a^{2N-4} \left(A_{2}(a)-A_{1}(a)\right)^{2}}{4 a_{0}^{2N-4}\left(\sqrt{A_{2}(a_{0})}-\sqrt{A_{1}(a_{0})}\right)^2},
\label{potencial}
\end{equation}
can be interpreted as an effective potential. This potential satisfies that $V(a_0)=0$ and, by using Eq. (\ref{CondEstatico}), $V'(a_0)=0$ as well. The second derivative of the Eq. (\ref{potencial}) evaluated at the radius $a_0$ results
\begin{eqnarray}
V''(a_0)&=& -\frac{(N-2)(2N-3) \left(\sqrt{A_{2}(a_{0})}-\sqrt{A_{1}(a_{0}})\right)^{2}}{2a_{0}^2}\nonumber \\  
&&-\frac{(N-2)(2N-5)\left(\sqrt{A_{1}(a_{0})}+\sqrt{A_{2}(a_{0}})\right)^{2}}{2a_{0}^2}-\frac{\left(A_{2}'(a_{0})-A_{1}'(a_{0})\right)^2}{2 \left(\sqrt{A_{2}(a_{0})}-\sqrt{A_{1}(a_{0})}\right)^{2}}\nonumber \\
&&-\frac{2(N-2)\left(\sqrt{A_{1}(a_{0})}+\sqrt{A_{2}(a_{0}})\right)^{2}\left(A_{2}'(a_{0})-A_{1}'(a_{0})\right)}{a_{0}\left(A_2(a_0)-A_1(a_0)\right)}  \nonumber \\
&&+\frac{A_{1}''(a_{0})+A_{2}''(a_{0})}{2}
-\frac{\left(\sqrt{A_{1}(a_{0})}+\sqrt{A_{2}(a_{0})}\right)^{2}\left(A_{2}''(a_{0})-A_{1}''(a_{0})\right)}{2\left(A_{2}(a_{0})-A_{1}(a_{0})\right)}.
\label{potencial2der}
\end{eqnarray}
If $V''(a_0)>0$, we can guarantee that the static configuration with radius $a_0$ is stable under radial perturbations.

\section{Examples}\label{examples} 

We present two different examples, in which we adopt the geometry (\ref{metric-N-sphe}) with the metric function given by Eq. (\ref{AmetricNd}), in order to describe both the inner $\mathcal{M}_1$ and the external $\mathcal{M}_2$ regions of the spacetime $\mathcal{M}$. One corresponds to bubbles and the other to thin shells surrounding black holes. The radius of the shell of matter in these constructions is determined by the static solution $a_0$ of Eq. (\ref{CondEstatico}). The configuration is stable under radial perturbations when the second derivative of the effective potential, given by Eq. (\ref{potencial2der}), is positive. The results, chosen among the most representative ones, are shown graphically in Figs. \ref{fig1}, \ref{fig2}, \ref{fig3}, and \ref{fig4}. In all these plots, the solid lines represent the stable solutions while the dashed lines the unstable ones. The meshed regions indicate zones where the matter satisfies the WEC condition, therefore the solution is made of normal matter, while the gray regions have no physical meaning. We limit the presentation of the results to $N=4$ and $N=5$, since the bigger the dimension of the spacetime, the larger the scales needed, showing no change in the overall behavior of the static solutions. We are interested in the higher dimensional case, the scenario with $N=4$ was previously studied \cite{tsFR} and it is shown here only for comparison. 

\subsection{Bubbles}

\begin{figure}[t!]
\centering
\includegraphics[width=0.8\textwidth]{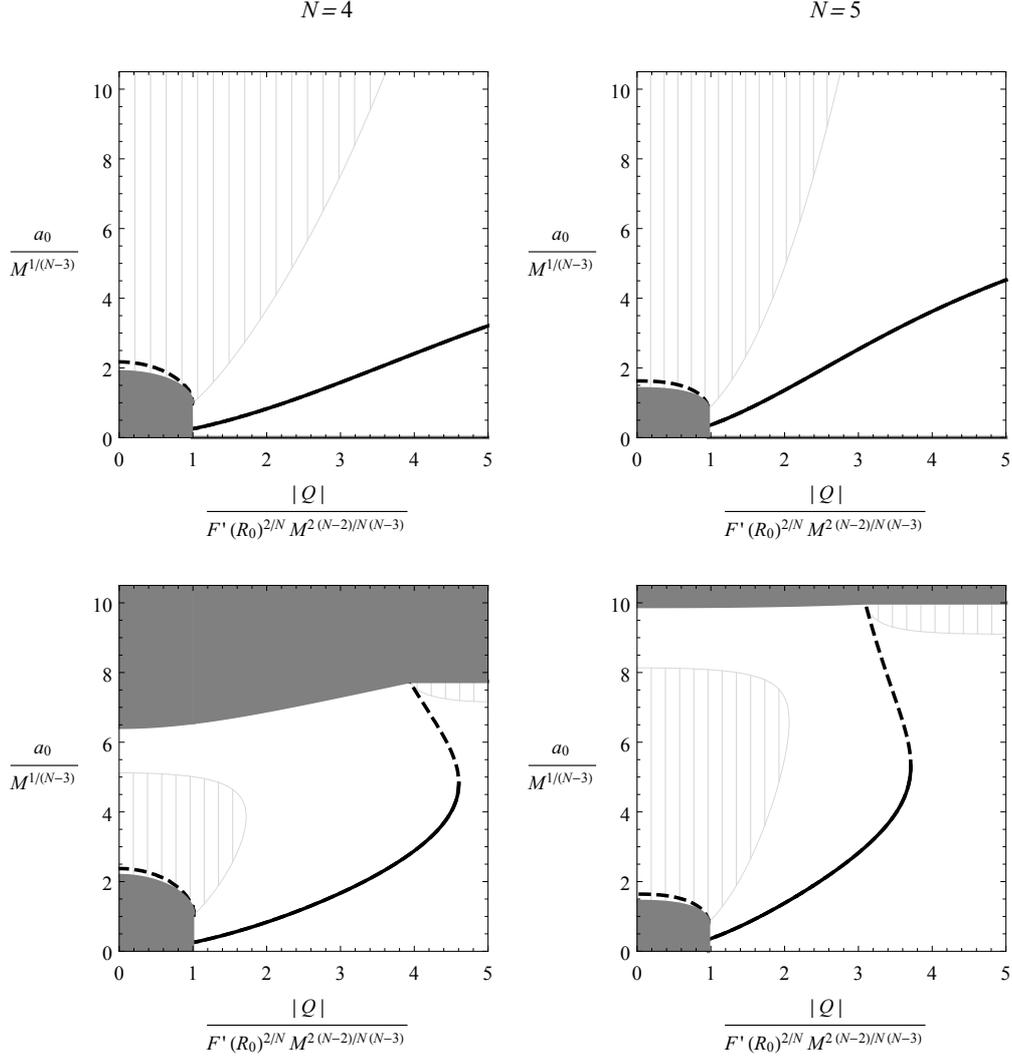}
\caption{Bubbles for $R_1=R_2=R_0$. The dimensionality $N$ is indicated in each column, stable configurations are displayed by solid lines and unstable ones by dashed lines, while in the meshed regions the matter is normal. The first row corresponds to the plots with $R_0 M^{2/(N-3)}=-0.2$ and the second row with $R_0 M^{2/(N-3)}=0.2$.}
\label{fig1}
\end{figure}

\begin{figure}
\centering
\includegraphics[width=0.8\textwidth]{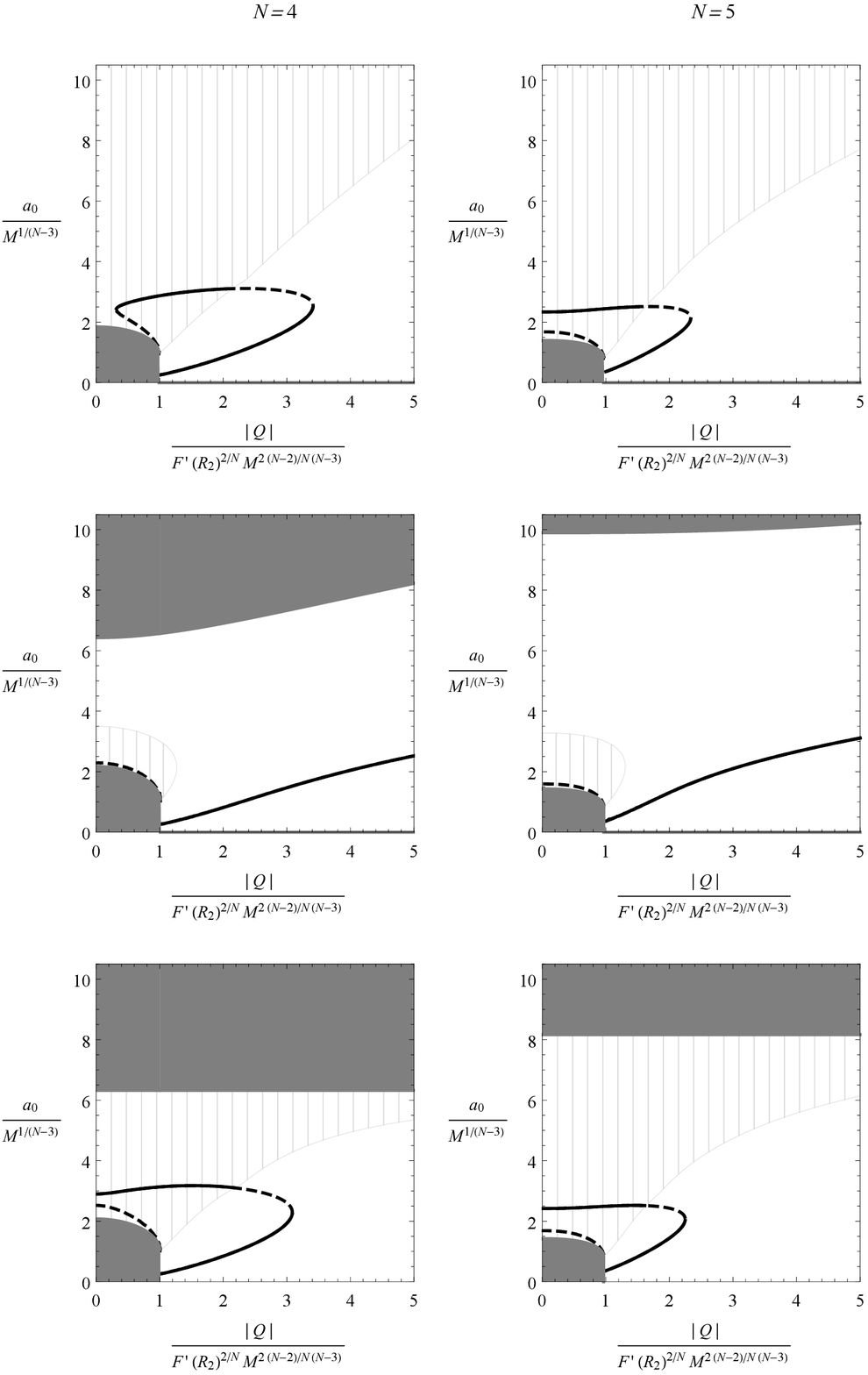}
\caption{Bubbles for $R_1\neq R_2$. The meaning of $N$, the solid and the dashed lines, and the meshed regions are the same as in Fig. \ref{fig1}. The first row shows the plots for $R_1 M^{2/(N-3)}=-0.1$ and $R_2 M^{2/(N-3)}=-0.3$, the second row for $R_1 M^{2/(N-3)}=-0.2$ and $R_2 M^{2/(N-3)}=0.2$, and the third row for $R_1 M^{2/(N-3)}=0.3$ and $R_2 M^{2/(N-3)}=0.1$.}
\label{fig2}
\end{figure}

In the construction of bubbles, we use the metric function given by Eq. (\ref{AmetricNd}), where we take $M_1=0$ and $Q_1=0$ for the inner zone $\mathcal{M}_1$, that is,
\begin{equation}
A_1(r)= 1 -\frac{R_1 r^2}{(N-1) N},
\label{metricinner}
\end{equation}
while we adopt $M_2=M$, $Q_2=Q$, and $\alpha =1$ for the external one $\mathcal{M}_2$, so that
\begin{equation}
A_2(r)= 1-\frac{2 M}{r^{N-3}}+\frac{2^{N/4}|Q|^{N/2}}{2F'(R_2)r^{N-2} } -\frac{R_2 r^2}{(N-1) N}.
\label{metricexternal}
\end{equation}
In this way, a vacuum region with scalar curvature $R_1$ is joined by a thin shell to an external one with scalar curvature $R_2$, mass $M$, and charge $Q$. The inner geometry has a cosmological horizon if $R_1>0$, otherwise it has no horizons. In the case of the external geometry, there is a critical value of charge $Q_c$: when $|Q|\le Q_c$  it has an event horizon and if $|Q|>Q_c$ the singularity at the origin is naked; for $R_2>0$, there is also a cosmological horizon. The radius $a_0$ should be large enough to avoid the presence of the event horizon when $|Q| \le Q_c$ and it also has to be smaller than any of the cosmological horizon radii when $R_1>0$ or $R_2>0$. There is a null electric field in the inner region and a radial electric field $E(r)=F_{tr}=Q/r^2$ in the outer one; this field has a jump $[F_{tr}]=Q/a_0^2$ at $\Sigma$, so we can interpret $Q$ as the charge of the shell. Then, the spacetime consists of a vacuum region surrounded by a charged thin shell, embedded in a region with a non-null electromagnetic field and de Sitter ($R_2>0$) or anti-de Sitter ($R_2<0$) asymptotics. For this spacetime, we present the results in  Figs. \ref{fig1} and \ref{fig2}. In them, the columns display the dimension of the spacetime, $N=4$ and $N=5$, respectively, while the rows correspond to different values of the constant scalar curvature $R$. The value of the mass $M$ establishes the scale of length and charge, so all quantities have been adimensionalized with $M$. 

In the case that both regions have the same scalar curvature at the sides of the shell $R_1=R_2=R_0$, shown in Fig. \ref{fig1}, the first row has $R_0 M^{2/(N-3)}=-0.2$, while the second one has $R_0 M^{2/(N-3)}=0.2$. From these plots, for fixed $N$ and $M$, we can say that
\begin{itemize}
\item When $R_0<0$, we find one solution for $|Q| \le Q_c$, which is unstable and made of normal matter. For $|Q| > Q_c$, the only solution is stable and composed by exotic matter.
\item When $R_0>0$, there is one solution for $|Q| \le Q_c$, which is unstable and made of normal matter. For $|Q| > Q_c$, the solution becomes stable and composed by exotic matter; a second solution also appears for a short range of $|Q|$, unstable and made of exotic matter, which can have a large radius.
\end{itemize}
The qualitative behavior of the solutions changes with the sign of the scalar curvature $R_0$ but it remains the same while increasing the dimension $N$ of the spacetime.

Within quadratic theories, i.e., $F(R)= R-2\Lambda + \gamma R^2$, the condition of continuity of the scalar curvature can be relaxed, so we can build a bubble with different constant values $R_1\neq R_2$ across the shell. The corresponding results are shown in Fig. \ref{fig2}, in which we have taken $\gamma/M= 0.1$. In the first row, we have the combination of values of $R_1 M^{2/(N-3)}=-0.1$ and $R_2 M^{2/(N-3)}=-0.3$, in the second row,  $R_1 M^{2/(N-3)}=-0.2$ and $R_2 M^{2/(N-3)}=0.2$, and in the third row, $R_1 M^{2/(N-3)}=0.3$ and $R_2 M^{2/(N-3)}=0.1$.  The behavior of the solutions, for fixed $N$ and $M$, can be summarized as:
\begin{itemize}
\item When $R_1<R_2$, for $|Q| \le Q_c$, we find one unstable solution made of normal matter, while for $|Q| > Q_c$, the only solution is stable and composed by exotic matter.
\item When $R_1>R_2$, for $|Q| \le Q_c$, we find two solutions made of normal matter; the larger one is stable, while the other is unstable. For $|Q| > Q_c$, we obtain two solutions again: the smaller one is always made of exotic matter and stable, while the other, depending on the value of $|Q|$, is stable and has normal matter, or it is unstable and has exotic matter. Given adequate values of the parameters, we can find two non-charged solutions composed by normal matter, the larger one is stable, while the other is unstable.
\end{itemize}
It is worth noticing that the behavior of the solutions strongly depends on the relative values of the scalar curvature. Modifying the dimension of the spacetime only affects the scale.

\subsection{Thin shells surrounding black holes}

\begin{figure}[t!]
\centering
\includegraphics[width=0.8\textwidth]{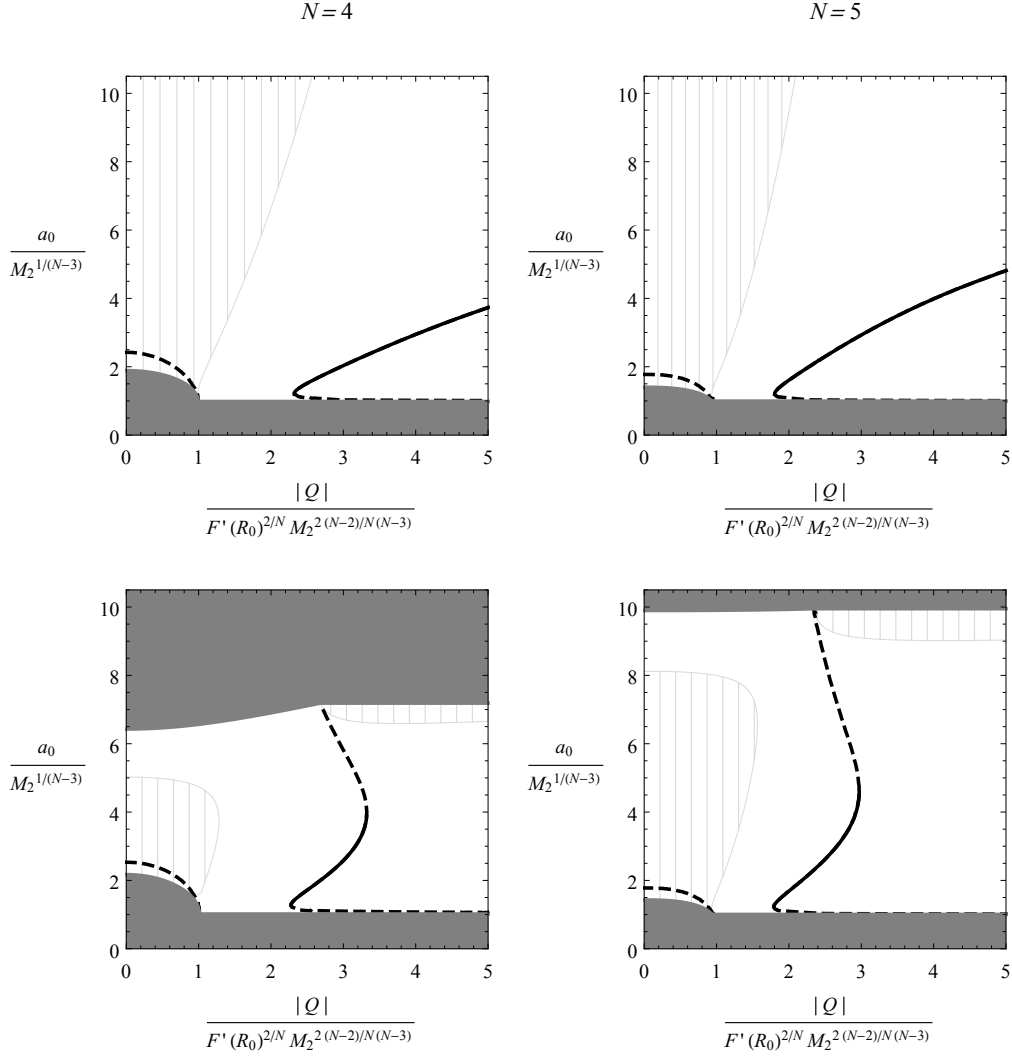}
\caption{Thin shell surrounding a black hole for $R_1=R_2=R_0$. The meaning of $N$, the solid and the dashed lines, and the meshed regions are the same as in Fig. \ref{fig1}.  The first row corresponds to the plots with $R_0 M_2^{2/(N-3)}=-0.2$ and the second row with $R_0 M_2^{2/(N-3)}=0.2$.}
\label{fig3}
\end{figure}

\begin{figure}
\centering
\includegraphics[width=0.8\textwidth]{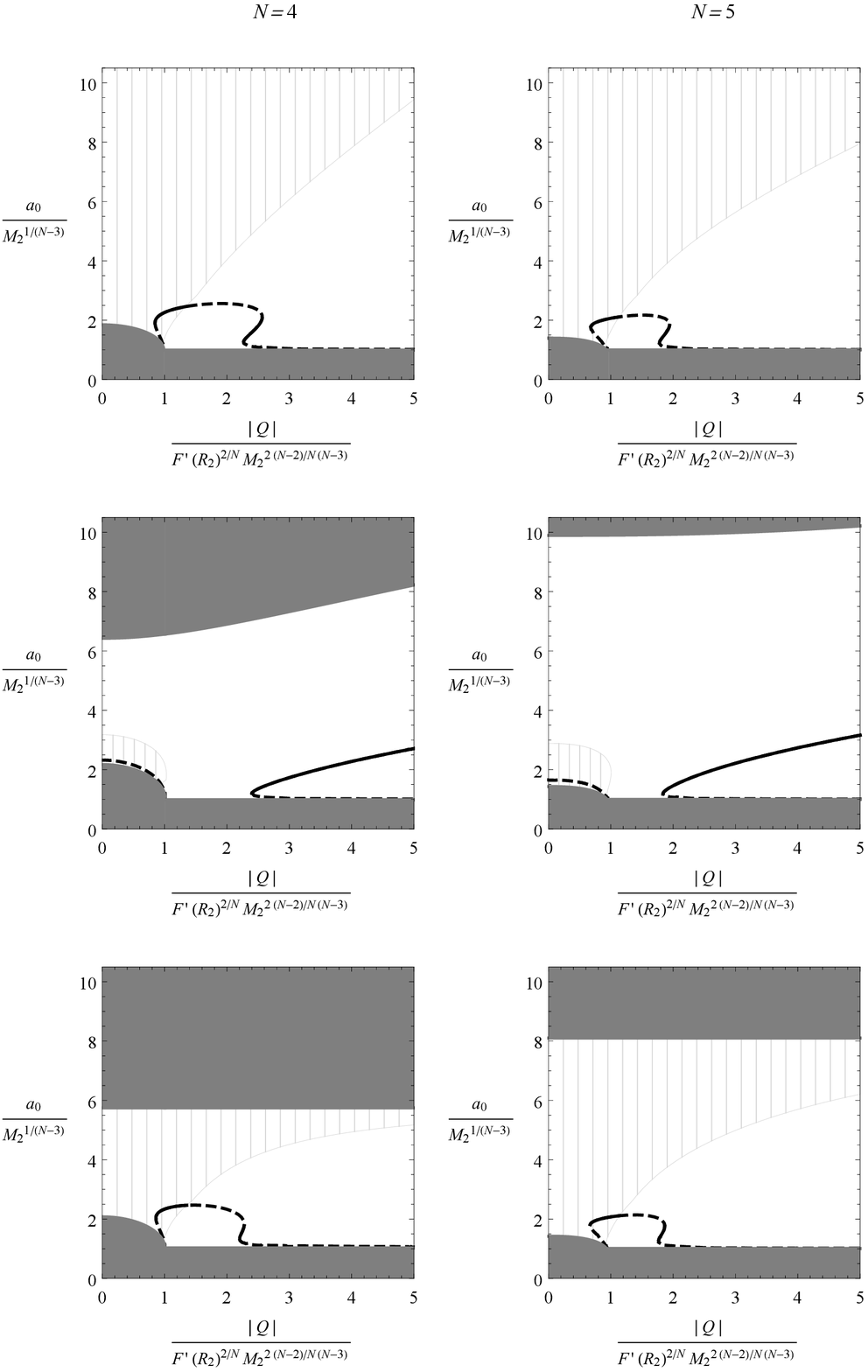}
\caption{Thin shell surrounding a black hole for $R_1\neq R_2$.  
The meaning of $N$, the solid and the dashed lines, and the meshed regions are the same as in Fig. \ref{fig1}. The first row shows the plots for $R_1 M_2^{2/(N-3)}=-0.1$ and $R_2 M_2^{2/(N-3)}=-0.3$, the second row for $R_1 M_2^{2/(N-3)}=-0.2$ and $R_2 M_2^{2/(N-3)}=0.2$, and the third row for $R_1 M_2^{2/(N-3)}=0.3$ and $R_2 M_2^{2/(N-3)}=0.1$.}
\label{fig4}
\end{figure}

We now construct thin shells surrounding black holes, with constant values of the scalar curvature $R_1$ and $R_2$ at the sides of the hypersurface $\Sigma$. We proceed in the same way as described in the previous subsection, with the only difference being that the inner zone $\mathcal{M}_1$  in our construction has non-null mass, i.e., $M_1\neq 0$. Its metric function reads
\begin{equation}
A_1(r)= 1-\frac{2 M_1}{r^{N-3}} -\frac{R_1 r^2}{(N-1) N}
\label{metricinnerMdif0}
\end{equation}
while the metric of the external region $\mathcal{M}_2$ is given by Eq. (\ref{metricexternal}) with $M$ replaced by $M_2$. In this construction, a  non-charged black hole with mass $M_1$ is surrounded by a thin shell connecting to an outer region with mass $M_2$ and charge $Q$. We take the radius $a_0$ larger than the event horizon radius of the inner region, so the black hole is always present, and when $ R_1> 0 $ also smaller than the cosmological horizon radius for this geometry. On the other hand, when $|Q| \le Q_c$, the radius $a_0$ should be large enough to avoid the presence of the event horizon of the geometry used for the outer region; when $ R_2 > 0 $, it also has to be smaller than the cosmological horizon radius of this zone.  As in the previous example, the electric field is null in the inner region and is given by $E(r)=F_{tr}=Q/r^2$ in the outer one; having a jump $[F_{tr}]=Q/a_0^2$ at $\Sigma$, which allows interpreting $Q$ as the charge of the shell. The whole spacetime consists of a non-charged black hole surrounded by a charged shell, having de Sitter ($R_2>0$) or anti-de Sitter ($R_2<0$) asymptotics. The results are shown in Figs. \ref{fig3} and \ref{fig4}; in the plots, the magnitudes have been adimensionalized with $M_2$ and hold the relation $M_1/M_2= 0.5$. 

The case with the same value of the scalar curvature $R_1=R_2=R_0$ at both sides of $\Sigma$ is shown in Fig. \ref{fig3}.  The first row corresponds to $R_0 M_2^{2/(N-3)}=-0.2$, while the second row to $R_0 M_2^{2/(N-3)}=0.2$. For given $N$ and $M_2$, we can see that:
\begin{itemize}
\item When $R_0<0$, we find one solution for $|Q| \le Q_c$, which is unstable and made of normal matter, and two solutions that appear from a certain value of $|Q|$ larger than $Q_c$, composed by exotic matter; in this case, the larger one is stable, while the other is unstable.
\item When $R_0>0$, there is one solution for $|Q| \le Q_c$, which is unstable and made of normal matter. For a short range of values of $|Q|$ larger than $Q_c$, there are three solutions made of exotic matter: the largest one is unstable and it can have a large radius, the middle one is stable, and the smaller one, which remains as the only solution for large values of $|Q|$, is unstable. 
\end{itemize}
The behavior of the solutions depends on the sign of the scalar curvature and it remains basically the same despite the dimension, which only alters the scale.

The scenario with $R_1 \neq R_2$ is shown in Fig. \ref{fig4}, in which we have adopted $\gamma /M_2 =0.1$. In the first row, we take the combination of values $R_1 M_2^{2/(N-3)}=-0.1$ and $R_2 M_2^{2/(N-3)}=-0.3$, in the second row,  $R_1 M_2^{2/(N-3)}=-0.2$ and $R_2 M_2^{2/(N-3)}=0.2$, and in the third row, $R_1 M_2^{2/(N-3)}=0.3$ and $R_2 M_2^{2/(N-3)}=0.1$. For given $N$ and $M_2$, the main features are:
\begin{itemize}
\item When $R_1<R_2$, there is only one solution for $|Q| \le Q_c$, unstable and made of normal matter. From  a certain value of $|Q|$ larger than $Q_c$, we find two solutions, the larger one is stable, while the other is unstable, both are composed by exotic matter.
\item When $R_1>R_2$, from a certain value of $|Q|$ smaller than $Q_c$, there are two solutions, the larger one is stable and the other is unstable, both are made of normal matter. As $|Q|$ increases only one solution remains, which at first is stable and composed by normal matter, and then becomes unstable and with exotic matter. For a short range of $|Q|$, we can find three solutions, the middle one is the only stable, all are made of exotic matter. Finally, for large $|Q|$, there is only one solution, which is unstable and composed by exotic matter.
\end{itemize}
The characteristics of the solutions mainly depend on the relationship between the different scalar curvatures. Again, we find that the dimension only affects the scale of the solutions.

\section{Summary}\label{summary}

In this work, we have found a generalized black hole solution with spherical symmetry for $N$-dimensional $F(R)$ gravity coupled to a conformally invariant Maxwell field, with constant scalar curvature $R$. It provides a generalization of the one corresponding to general relativity as a particular case. We have compared our solution with others existing in the literature. We have also constructed a family of spacetimes with a spherically symmetric thin shells in $N$-dimensional $F(R)$ gravity with constant $R$ and we have studied the stability of the static configurations under radial perturbations. In order to avoid the presence of ghosts, we have always worked with $F'(R)>0$.

We have used our generalized solution in the two given examples of our formalism, one representing bubbles and the other thin layers of matter surrounding black holes. These spacetimes, with charge $Q$ and a radial electric field, have been built within general $F(R)$ theory, which forces us to work with the same value of the constant scalar curvature $R_0$ across the shell, and also in quadratic $F(R)$, which allows different constant values $R_1$ and $R_2$ of the scalar curvature. In each example, we have found the expressions for the energy density and the pressure at the shell, and the corresponding equation of state. In quadratic $F(R)$, we have also found the extra contributions at the shell, present when $R_1 \neq R_2$. We have analyzed the stability of the configurations for the different combinations of the parameters. 

For bubbles, in the case with the same value $R_0$ at both sides of the shell, we have obtained that stable configurations are possible, but composed by exotic matter. The behavior of these configurations mainly depends on the sign of $R_0$. For quadratic $F(R)$ with $R_1 \neq R_2$, we have found stable solutions, which are made of normal matter only when $R_1>R_2$. In this case, given a particular set of parameters, there exist stable solutions even without charge and composed by normal matter. The relationship between $R_1$ and $R_2$ is what determines the behavior of the solutions. The dimension of the spacetime modifies the scale without affecting the main characteristics of the solutions. 

In the case of thin layers of matter surrounding black holes, we have found stable configurations, but made of exotic matter for the same value $R_0$ across the shell, with their characteristics depending on the sign of  $R_0$. In quadratic $F(R)$ with $R_1 \neq R_2$, the relationship between $R_1$ and $R_2$ is what determines the behavior of the solutions. We have found stable solutions, which are composed by normal matter only in the case that $R_1>R_2$. Once again, the dimension of the spacetime only changes the scale of the solutions.

\section*{Acknowledgments}

This work has been supported by CONICET and Universidad de Buenos Aires.




\begin{thebibliography}{99}

\bibitem{sofa} T.P. Sotiriou and V. Faraoni, Rev. Mod. Phys.  \textbf{82}, 451 (2010); A. De Felice and S. Tsujikawa, Living Rev. Relativity \textbf{13}, 3 (2010); S. Nojiri, S.D. Odintsov, and  V.K. Oikonomou, Phys. Rep. \textbf{692}, 1 (2017).

\bibitem{bhfr1}  T. Multam\"aki and I. Vilja, Phys. Rev. D \textbf{74}, 064022 (2006); S. Capozziello, A. Stabile, and A. Troisi, Class. Quantum Gravity \textbf{25}, 085004 (2008).

\bibitem{bhfr2} A. de la Cruz-Dombriz, A. Dobado, and A.L. Maroto, Phys. Rev. D  \textbf{80}, 124011 (2009); \textbf{83}, 029903(E) (2011); T. Moon, Y.S. Myung, and E.J. Son, Gen. Relativ. Gravit. \textbf{43}, 3079 (2011).

\bibitem{bhfr3} L. Sebastiani and S. Zerbini, Eur. Phys. J. C \textbf{71}, 1591 (2011); S. Habib Mazharimousavi, M. Halilsoy, and T. Tahamtan, Eur. Phys. J. C \textbf{72}, 1851 (2012);   P. Ca\~nate, L.G. Jaime, and M. Salgado, Class. Quantum Gravity \textbf{33}, 155005 (2016); P. Cañate, Class. Quantum Gravity \textbf{35}, 025018 (2018); E. Elizalde, G.G.L. Nashed, S. Nojiri, and S.D. Odintsov, Eur. Phys. J. C \textbf{80}, 109 (2020); G.G.L. Nashed and S. Nojiri, Phys. Rev. D \textbf{102}, 124022 (2020); G.G.L. Nashed and S. Nojiri, Phys. Lett. B \textbf{820}, 136475 (2021).

\bibitem{whfr1} A. DeBenedictis and D. Horvat, Gen. Relativ. Gravit. \textbf{44}, 2711 (2012); T. Harko, F.S.N. Lobo, M.K. Mak, and S.V. Sushkov, Phys. Rev. D \textbf{87}, 067504 (2013).

\bibitem{whfr2} J.L. Rosa, J.P.S. Lemos, and F.S.N. Lobo, Phys. Rev. D \textbf{98}, 064054 (2018); H. Golchin and  M.R. Mehdizadeh, Eur. Phys. J. C  \textbf{79}, 777 (2019); F.S.N. Lobo, G.J. Olmo, E. Orazi, D. Rubiera-Garcia, and A. Rustam, Phys. Rev. D \textbf{102}, 104012 (2020).

\bibitem{BHRGndim1} M. Hassa\"ine and C. Mart\'inez, Phys. Rev. D \textbf{75}, 027502  (2007).

\bibitem{BHRGndim2} M. Hassa\"ine and C. Mart\'inez, Class. Quantum Gravity \textbf{25}, 195023 (2008).

\bibitem{BHRGndim3} S. Habib Mazharimousavi, Class. Quantum Gravity \textbf{37}, 197001 (2020).

\bibitem{BHRGndim4} D. Koko\v{s}ka and M. Ortaggio, Phys. Rev. D \textbf{104}, 124051 (2021).

\bibitem{BHFRndim1} S.H. Hendi, Phys. Lett. B \textbf{690}, 220 (2010);
S.H. Hendi, B. Eslam Panah, and S.M. Mousavi, Gen. Relativ. Gravit. \textbf{44}, 835 (2012).

\bibitem{BHFRndim2} A. Sheykhi, Phys. Rev. D \textbf{86}, 024013  (2012).

\bibitem{BHFRndim3} Z.Y. Tang, B. Wang, and E. Papantonopoulos,  Eur. Phys. J. C  \textbf{81}, 346 (2021).

\bibitem{BTZ92}  M. Ba\~nados, C. Teitelboim, and J. Zanelli, Phys. Rev. Lett. \textbf{69}, 1849 (1992).

\bibitem{BHRG3dim1} M. Cataldo, N. Cruz, S. del Campo, and A. Garc\'ia, Phys. Lett. B, \textbf{484}, 154 (2000).

\bibitem{BHRG3dim2} O. Gurtug, S. Habib Mazharimousavi, and M. Halilsoy, Phys. Rev.  D \textbf{85}, 104004 (2012).

\bibitem{BHFR3dim} S.H. Hendi, B. Eslam Panah, and R. Saffari, Int. J. Mod. Phys. D \textbf{23}, 1450088 (2014).

\bibitem{daris} G. Darmois, M\'{e}morial des Sciences Math\'{e}matiques, Fascicule XXV, Chap. V (Gauthier-Villars, Paris, 1927); W. Israel, Nuovo Cimento B \textbf{44}, 1 (1966); \textbf{48}, 463(E) (1967).

\bibitem{gravstar} M. Visser and D.L. Wiltshire, Class. Quantum Gravity \textbf{21}, 1135 (2004);F. S. N. Lobo and A. V. B. Arellano, Class. Quantum Gravity \textbf{24}, 1069 (2007); P. Martin-Moruno, N. Montelongo Garcia, F.S.N. Lobo, and M. Visser, J. Cosmol. Astropart. Phys.  \textbf{03}, 034 (2012).

\bibitem{whrg} E. Poisson and M. Visser, Phys. Rev. D \textbf{52}, 7318 (1995); E.F. Eiroa and G.E. Romero, Gen. Relativ. Gravit. \textbf{36}, 651 (2004); E.F. Eiroa, Phys. Rev. D \textbf{78}, 024018 (2008); N. Montelongo Garcia, F.S.N. Lobo, and M. Visser, Phys. Rev. D \textbf{86}, 044026 (2012); S.D. Forghani, S. Habib Mazharimousavi, and M. Halilsoy, Eur. Phys. J. C \textbf{78}, 469 (2018); T. Berry, F.S.N. Lobo, A. Simpson, and M. Visser, Phys. Rev. D \textbf{102}, 064054 (2020).

\bibitem{shrg} P.R. Brady, J. Louko and E. Poisson, Phys. Rev. D \textbf{44}, 1891 (1991); M. Ishak and K. Lake, Phys. Rev. D \textbf{65}, 044011 (2002); F.S.N. Lobo and P. Crawford, Class. Quantum Gravity \textbf{22}, 4869 (2005); E.F. Eiroa and C. Simeone, Phys. Rev. D \textbf{83}, 104009 (2011);  M. Sharif and S. Iftikhar, Astrophys. Space Sci. \textbf{356}, 89 (2015).

\bibitem{GRddim} G.A.S. Dias and J.P.S. Lemos, Phys. Rev. D \textbf{82}, 084023 (2010); F. Rahaman, M. Kalam, and S. Chakraborty , Gen. Relativ. Gravit. \textbf{38}, 1687 (2006),A. Banerjee, K. Jusufi, and S. Bahamonde, Grav. Cosmol. \textbf{24}, 71 (2018);  
E.F. Eiroa and C. Simeone, Int. J. Mod. Phys. D \textbf{21}, 1250033 (2012).

\bibitem{dss} N. Deruelle, M. Sasaki, and Y. Sendouda, Prog. Theor. Phys. \textbf{119}, 237 (2008).

\bibitem{js} J.M.M. Senovilla, Phys. Rev. D \textbf{88}, 064015 (2013); J.M.M.  Senovilla, Class. Quantum Gravity \textbf{31}, 072002 (2014); J.M.M.  Senovilla, J. Phys. Conf. Ser. \textbf{600}, 012004 (2015); B. Reina, J.M.M. Senovilla, and R. Vera, Class. Quantum Gravity \textbf{33}, 105008 (2016).

\bibitem{TSWHinFR} E.F. Eiroa and G. Figueroa-Aguirre, Eur. Phys. J. C \textbf{76}, 132 (2016);  E.F. Eiroa and G. Figueroa-Aguirre, Phys. Rev. D \textbf{94}, 044016 (2016); M. Zaeem-ul-Haq Bhatti, A. Anwar, and S. Ashraf, Mod. Phys. Lett. A \textbf{32}, 1750111 (2017); S. Habib Mazharimousavi, Eur. Phys. J. C \textbf{78}, 612 (2018); S. Habib Mazharimousavi, M. Halilsoy, and K. Kianfar, Eur. Phys. J. Plus \textbf{135}, 440 (2020).

\bibitem{tsFR} E.F. Eiroa, G. Figueroa-Aguirre, and J.M.M. Senovilla, Phys. Rev. D \textbf{95}, 124021 (2017), E.F. Eiroa and G. Figueroa-Aguirre, Eur. Phys. J. C \textbf{78}, 54 (2018); E.F. Eiroa and G. Figueroa-Aguirre, Eur. Phys. J. C \textbf{79}, 171 (2019).

\bibitem{tsFR3d} E. F. Eiroa and G. Figueroa-Aguirre, Phys. Rev.  D \textbf{103}, 044011 (2021); C. Bejarano, E. F. Eiroa, and G. Figueroa-Aguirre, Eur. Phys. J. C \textbf{81}, 668 (2021).

\bibitem{bronnikov} K.A. Bronnikov, M.V. Skvortsova, and A.A. Starobinsky, Grav. Cosmol. \textbf{16}, 216 (2010).

\end{thebibliography}
\end{document}